\documentclass[conference]{IEEEtran}

\IEEEoverridecommandlockouts
\usepackage{cite}
\usepackage{amsmath,amssymb,amsfonts}
\usepackage{algorithmic}
\usepackage{graphicx}
\usepackage{textcomp}
\usepackage{xcolor}

\usepackage{cite}
\usepackage{amsmath,amssymb,amsfonts}
\usepackage{algorithmic}
\usepackage{graphicx}
\usepackage{textcomp}
\usepackage{subfigure}
\usepackage{algorithm}
\usepackage{algorithmic}
\usepackage{xcolor}%定义了一些颜色  
\usepackage{colortbl,booktabs}%第二个包定义了几个*rule
\usepackage{cite}
\usepackage{amsmath}
\usepackage{threeparttable}  
\usepackage{epstopdf}
\usepackage{booktabs}  
\usepackage{diagbox}
\usepackage{multirow}
\usepackage{multicol}
\usepackage{caption}
\usepackage{stfloats}
\usepackage{pdfpages}
\usepackage{setspace}
\usepackage{fancyhdr}
\usepackage{bm}

\def\BibTeX{{\rm B\kern-.05em{\sc i\kern-.025em b}\kern-.08em
    T\kern-.1667em\lower.7ex\hbox{E}\kern-.125emX}}
\begin{document}

\title{A Robust Deep Learning Enabled Semantic Communication System for Text\\

}

\author{
	\IEEEauthorblockN{
	Xiang Peng\IEEEauthorrefmark{1},
	Zhijin Qin\IEEEauthorrefmark{2},
	Danlan Huang\IEEEauthorrefmark{1}\IEEEauthorrefmark{3}, 
	Xiaoming Tao\IEEEauthorrefmark{1}\IEEEauthorrefmark{3}\thanks{Xiaoming Tao is the corresponding author.},
	Jianhua Lu\IEEEauthorrefmark{1}\IEEEauthorrefmark{3},
	Guangyi Liu\IEEEauthorrefmark{4},
	Chengkang Pan\IEEEauthorrefmark{4}}

	\IEEEauthorblockA{\IEEEauthorrefmark{1}Department of Electronic Engineering, Tsinghua University, Beijing, China}
	\IEEEauthorblockA{\IEEEauthorrefmark{2}Queen Mary University of London, London, UK}
	\IEEEauthorblockA{\IEEEauthorrefmark{3}Beijing National Research Center for Information Science and Technology (BNRist), Beijing, China}
	\IEEEauthorblockA{\IEEEauthorrefmark{4}China Mobile Research Institute, China}
	\IEEEauthorblockA{Email: px21@mails.tsinghua.edu.cn, z.qin@qmul.ac.uk, \{huangdl, taoxm,  	lhh-dee\}@mail.tsinghua.edu.cn, \\
	\{liuguangyi, panchengkang\}@chinamobile.com}
	}

\maketitle

\begin{abstract}
With the advent of the 6G era, the concept of semantic communication has attracted increasing attention. Compared with conventional communication systems, semantic communication systems are not only affected by physical noise existing in the wireless communication environment, e.g., additional white Gaussian noise, but also by semantic noise due to the source and the nature of deep learning-based systems. In this paper, we elaborate on the mechanism of semantic noise. In particular, we categorize semantic noise into two categories: literal semantic noise and adversarial semantic noise. The former is caused by written errors or expression ambiguity, while the latter is caused by perturbations or attacks added to the embedding layer via the semantic channel. To prevent semantic noise from influencing semantic communication systems, we present a robust deep learning enabled semantic communication system (R-DeepSC) that leverages a calibrated self-attention mechanism and adversarial training to tackle semantic noise. Compared with baseline models that only consider physical noise for text transmission, the proposed R-DeepSC achieves remarkable performance in dealing with semantic noise under different signal-to-noise ratios.
\end{abstract}


\section{Introduction}
% \addtolength{\topmargin}{0.5in}
Distinct from conventional wireless communications, which focus on reducing transmission symbol errors, semantic communication targets to extract and interpret the meaning behind symbols accurately~\cite{semantic_noise}. Therefore, the optimization goal of semantic communication is to narrow the semantic gap between transmitted and received signals rather than lowing the bit error rate. Such a transmission goal determines that semantic communication is mainly applied for communications between agents, such as machine-to-machine communications or human-to-machine communications.

Recently developed semantic communication systems~\cite{lite_DeepSC, Federal_audio, DeepSC-S, DeepSC-VQA, danlan, I-DeepSC}, leverage the substantial power of deep neural networks (DNNs) in semantic extraction to understand the meaning of texts. DeepSC~\cite{lite_DeepSC} is a pioneer work on semantic communications that presents a novel and effective architecture for text semantic transmission. Works on semantic communications have also been extended to multiple tasks, such as speech transmission\cite{Federal_audio, DeepSC-S}, image transmission~\cite{danlan, I-DeepSC}, and visual question answering~\cite{DeepSC-VQA}. Most of these works take the impact of various kinds of physical channel noise into consideration and use a joint source-channel coding scheme to combat the influence of physical channel noise. 

However, besides physical channel noise, semantic noise can also affect the semantic communication system. The key factor that determines the performance of semantic communication is the fidelity of semantic information it extracts and processes, while semantic information may be disturbed by semantic noise. 

On the one hand, the original text may contain grammatical errors or slight literal modifications, such as deletions, replacement, order reversion, etc. These literal changes in texts will incur semantic distortion and obstruct the subsequent semantic understanding and interpretation~\cite{sememes}. For example, it is easy to mislead the model by adding a punctuation or character to a word of the text.~\cite{google}. 

On the other hand, due to the limited generalization ability, DNNs-based systems are vulnerable to malicious attacks. A slight perturbation added to input signals can render models misunderstand their semantics~\cite{adverarial_equ}. Consequently, the wrong decision will be made. For example, adding unperceived noise to a picture can deceive a classification model~\cite{2013Intriguing}. Analogously, adding noise to the embedding representation of a text may also affect the semantic extraction and result in a misunderstanding of the text~\cite{survey}.

Semantic noise could cause semantic ambiguity and make it hard for receivers to convey the underlying meaning of the transmitted text. Conventional communication systems are unable to handle such errors, because they are optimized at the symbol level. However, semantic communication systems are expected to overcome these disturbances and recover the original meaning from the modified text due to their semantic understanding ability.

In this paper, distinct from well-discussed physical noise on wireless channels, we explore different forms of semantic noise and establish a robust semantic communication system named R-DeepSC to effectively eliminate the impact of different kinds of semantic noise in text. To the best of our knowledge, this paper is the first to comprehensively explore semantic noise in text transmission. The detailed contributions of this paper are summarized as follows.

\begin{itemize}
    \item 
    %We categorize semantic noise in semantic communications as literal modifications to original texts or disturbances imposed on the text embedding, which are caused by sentence modifications or written errors, and adversarial attacks, respectively. To combat the semantic noise, we propose robust deep learning enabled semantic communication system, named R-DeepSC. 
    We categorize semantic noise in communications as literal modifications and adversarial noise. To combat the semantic noise, we propose a robust deep learning enabled semantic communication system named R-DeepSC. 
    
  \item  For the literal semantic noise, we tailor the transformer-based model and present a calibrated self-attention mechanism for error correction to ensure the semantic fidelity. 

   \item For the adversarial semantic noise, we adopt an adversarial training method to train the system. We experimentally verify the effectiveness of the R-DeepSC in resisting different forms of semantic noise. 

\end{itemize}

The remaining parts of this paper are organized as follows. Section \uppercase\expandafter{\romannumeral2} introduces various kinds of semantic noise and our anti-noise methods in detail. The experiment results are shown and discussed in Section \uppercase\expandafter{\romannumeral3}. Section \uppercase\expandafter{\romannumeral4} concludes the paper.
\begin{figure*}[htp]
	\centering
	\includegraphics[scale=0.7]{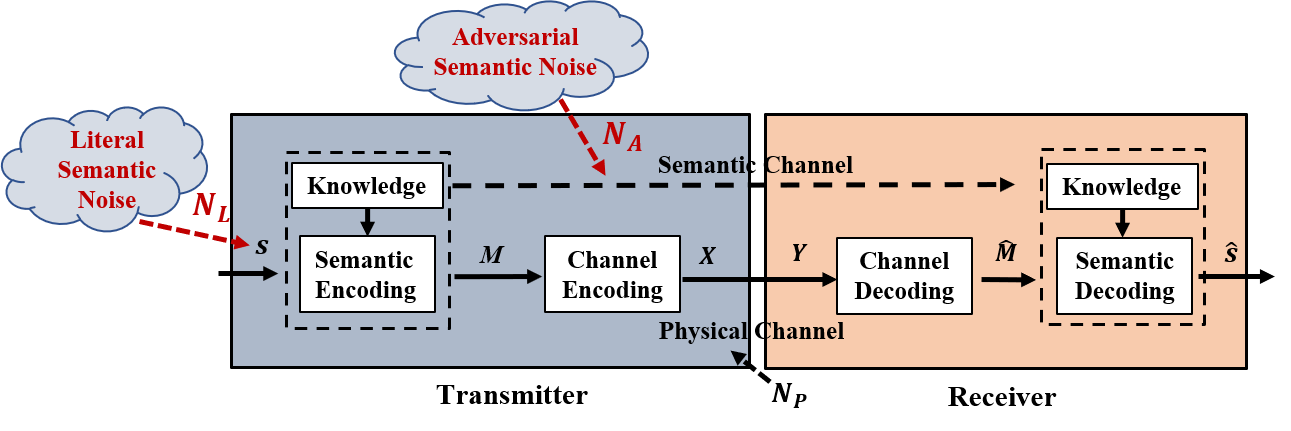}
	\caption{The semantic communication architecture and semantic noise.}
	\label{fig:1}
\end{figure*}

\section{Semantic Communication System Model}

In this section, we consider a semantic communication system with physical noise and semantic noise.

Fig.~\ref{fig:1} is the architecture of a semantic communication system. The transmitter conducts semantic encoding and channel encoding, while the receiver performs the corresponding decoding. Besides physical noise, $\bm{N_P}$, \textit{literal semantic noise}, $\bm{N_L}$, and \textit{adversarial semantic noise}, $\bm{N_A}$, could affect the considered semantic communication system.

The source text $\bm{S}$ can be affected by the literal semantic noise, $ \bm{N_{L}} $, which is defined as errors in $\bm{S}$, such as substitutions, deletions, etc. The literal semantic noise not only makes it difficult for humans to understand the underlying meaning of the text, but also incurs semantic distortions for semantic encoding. The text with literal semantic noise is given by $\mathcal{F} (\bm{S}, \bm{N_L}) $, where $ \mathcal{F}(\cdot) $ is a noise-adding function simulating the expression habits of users or 
vulnerable AI-assisted transmission environment, such as a speech recognition system. \textit{Literal semantic noise ratio} is defined as the proportion of erroneous words in a sentence.

We denote the input text of the system as $\bm{S}$, $\bm{S} = \{s_0, s_1, \cdots, s_L\}$, where $s_i$ is the i-the word. After $\bm{S}$ passes the one-hot encoder and the embedding layer, the embedding vector $\bm{X_{embed}}$ is represented as
\begin{equation}
\bm{X_{embed}} = E_{\bm{\gamma}}(O_{\bm{d}}(\mathcal{F}(\bm{S}, \bm{N_L))}),
\end{equation}
where $O_{\bm{d}}(\cdot)$ is the one-hot encoder according to dictionary $\bm{d}$ and $E_{\bm{\gamma}}(\cdot)$ is the embedding layer with the parameter set $\bm{\gamma}$. 

\begin{figure}[htp]
	\centering
	\includegraphics[scale=0.55]{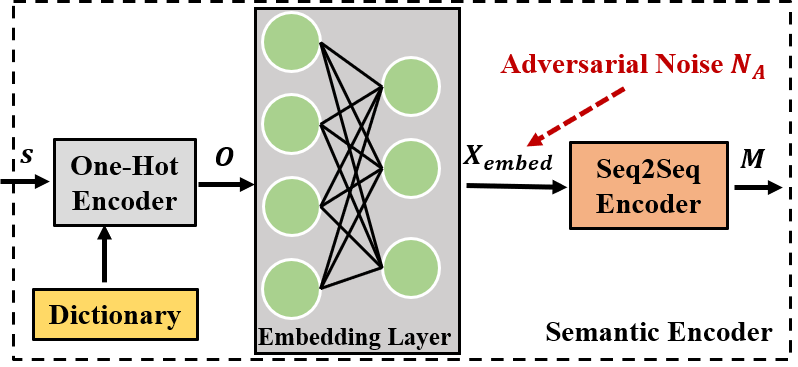}
	\caption{The developed transmitter of the semantic communication system.}
	\label{fig:FGM}
\end{figure}

The architecture of the transmitter is illustrated in Fig.\ref{fig:FGM}. During this process, the one-hot encoder can hardly be affected by undetectable interference due to its natural sparsity. Conversely, the adversarial semantic noise, $\bm{N_{A}}$, which is a slight perturbation added to the embedding vector, $\bm{X_{embed}}$, may cause semantic misunderstanding. By considering the adversarial semantic noise, the transmitted signal is given by
\begin{equation}
\bm{X} = C_{\bm{\varphi}}S_{\bm{\eta}}(\bm{X_{embed}} + \bm{N_A}),
\end{equation}
where $C_{\bm{\varphi}}(\cdot)$ is the channel encoder with the parameter set $\bm{\varphi}$, and $S_{\bm{\eta}}(\cdot)$ is the Seq2Seq encoder with the parameter set $\bm{\eta}$. The received signal, $\bm{Y}$, can be represented as
\begin{equation}
\bm{Y} = \bm{HX} + \bm{N_P},
\end{equation}
where $\bm{H}$ represents the fading channel and $\bm{N_P} \sim{\mathcal{CN} (0, {\sigma_n^ 2)}}$.
% where $H$ represents the Rayleigh fading channel with $\mathcal{CN} (0, 1)$ and $N_P \sim{\mathcal{CN} (0, {\sigma_n^ 2)}}$.

By utilizing the channel decoder and the semantic decoder, the received text $\bm{\hat{S}}$ can be represented as
\begin{equation}
\hat{\bm{S}} = C_{\bm{\zeta}} ^ {-1} (S_{\bm{\delta}} ^ {-1} (\bm{S})),
\end{equation}
where $C_{\bm{\zeta}}^{-1}(\cdot)$ is the channel decoder with the training parameter set $\bm{\zeta}$, and $S_{\bm{\delta}} ^ {-1}(\cdot)$ is the semantic decoder with the training parameter set $\bm{\delta}$.

The goal of this system is to minimize the semantic gap between transmitted text, $\bm{S}$, and reconstructed text, $\hat{\bm{S}}$. By representing the transmitter and receivers as neural networks, the loss function developed in DeepSC~\cite{lite_DeepSC} to train the system is given by
\begin{equation}
\mathcal{L}_{total}(\bm{S}, \bm{\hat{S}} ; \bm{\varphi}, \bm{\eta}, \bm{\zeta}, \bm{\delta}) = \mathcal{L}_{CE}(\bm{S}, \bm{\hat{S}}) + \alpha \cdot \mathcal{L}_{MI}(\bm{X}, \bm{Y}).
\end{equation}

\section{Anti-Semantic Noise Methods and Performance Metrics}

To combat semantic noise and maintain the semantic fidelity of the system, we propose a robust deep learning enabled semantic communication system named R-DeepSC. For the literal semantic noise, we develop a calibrated self-attention mechanism along with a novel loss function to eliminate literal errors. For the adversarial semantic noise, adversarial training is utilized to improve the robustness of the semantic communication system.

\subsection{Calibrated Self-Attention Mechanism}

Literal modifications can be operated at the character and word levels. By adopting a spelling-check method, character-level errors can be removed effectively~\cite{spell_correct}. Hence, we mainly focus on word-level literal semantic noise in this paper. Prior efforts have been made to solve the error correction problem from a data or model perspective~\cite{soft-mask, maskGEC}. A novel detect-correct framework was established to address the Chinese error-correction problem~\cite{soft-mask}. \cite{maskGEC} handled the grammatical error correction at the data level by leveraging a dynamic mask to generate error-correct examples for training. 

For semantic communications, to avoid errors from affecting semantic information, less attention should be paid to erroneous tokens when calculating semantic representation vectors. However, the self-attention mechanism is unable to realize this goal due to the absence of error probability information. To cope with this problem, a detection net is added to infer the error probability of each token.

\begin{figure}[htbp]
	\centering
	% 0.6
	\includegraphics[scale=0.50]{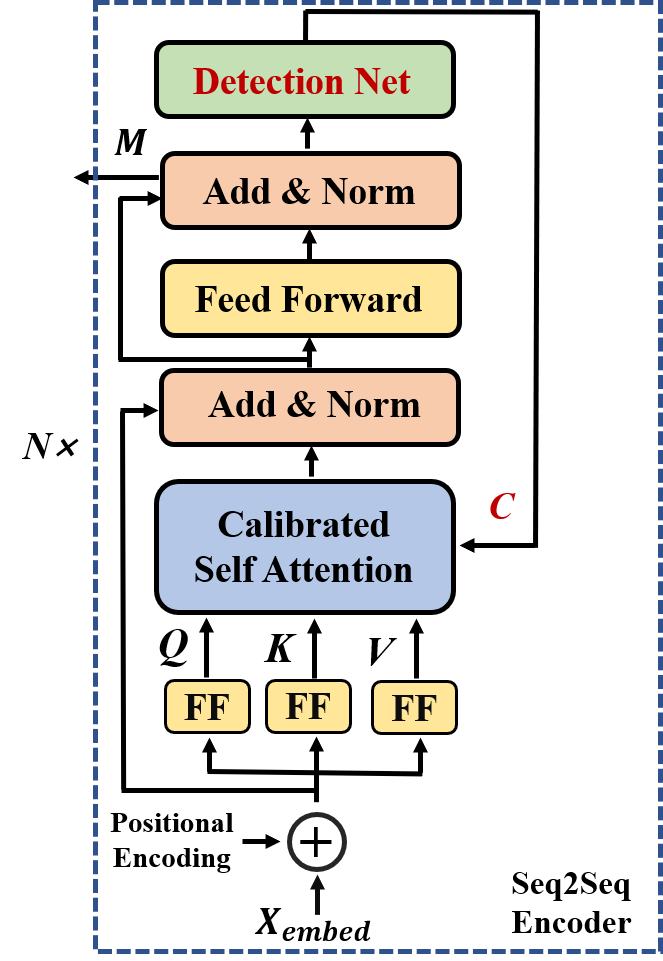}
	\caption{ The developed semantic encoder structure in R-DeepSC.}
	\label{fig: semantic encoder}
\end{figure}

The architecture of the semantic encoder developed in R-DeepSC is illustrated in Fig.~\ref{fig: semantic encoder}. The number of layers in the Transformer enabled Seq2Seq encoder is denoted as $N$. A detection net, which consists of a GRU and a linear layer, is added to the original semantic encoder of DeepSC. A calibration matrix, $ \bm{C} $, is obtained based on the output of the detection net. The attention score is calibrated by $C$ to ensure that more attention is devoted to unmistakable tokens.

The calibrated attention score can be represented by
\begin{equation}
\bm{C_{atten}} = SoftMax(\frac{\bm{Q} \cdot \bm{K ^ \mathrm{T}}}{\sqrt{d_k}} \cdot \bm{V} \times \bm{C}),
\end{equation}
where $ \times $ is element-wise product, $\bm{Q}$, $\bm{K}$, $\bm{V}$, $d_k$ is query, key, value and embedding dimension of the semantic encoder. 

To make the system robust to the literal semantic noise, we propose a new loss function to train the neural network of the developed R-DeepSC, which is given by
\begin{equation}
\begin{split}
\mathcal{L}_{total}(\bm{S}, \bm{\hat{S}} ; \bm{\varphi}, \bm{\eta}, \bm{\zeta}, \bm{\delta}) &= \mathcal{L}_{CE}(\bm{S}, \bm{\hat{S})} + \alpha \cdot \mathcal{L}_{MI}(\bm{X}, \bm{Y})\\
&+\beta\cdot \mathcal{L}_{BCE}(\bm{label}, \bm{P}), 
\end{split}
\end{equation}
where $\mathcal{L}_{CE}$ is the cross-entropy loss, $\mathcal{L}_{MI}$ is the mutual information, $\mathcal{L}_{BCE}$ is the binary cross-entropy loss, $\bm{P}$ is the error probability matrix of tokens that is predicted by the detection net, and $\bm{label}$ is the ground truth of the error probability. The proportions of $\mathcal{L}_{MI}$ and $\mathcal{L}_{BCE}$ in the loss function can be controlled by parameters $\alpha$ and $\beta$.

The loss function is utilized to optimize parameters, including $ \bm{\varphi}, \bm{\eta}, \bm{\zeta}, \bm{\delta}$. In this loss function, $\mathcal{L}_{CE}$ aims to make the transmitted text, $\bm{S}$, and the received text, $\hat{\bm{S}}$, as similar as possible, while the $\mathcal{L}_{MI}$ maximizes the channel capacity by maximizing the mutual information between the transmitted signal, $\bm{T_X}$, and the received signal, $\bm{R_X}$. $\mathcal{L}_{BCE}$ loss is applied to train the system for error probability prediction, which is an input of calibrated self-attention. 

\subsection{Adversarial Training}

For the adversarial semantic noise, adversarial training methods, such as fast gradient sign method (FGSM)~\cite{FGSM} and fast gradient method (FGM)~\cite{FGM}, were applied to eliminate its inference. \cite{semantic_noise_image} has discussed the processing of adversarial noise in semantic communications for images. However, these efforts have not yet been utilized to improve the robustness of semantic communication systems for text transmission. 

We take the advantage of adversarial training, which is able to productively improve the robustness of deep learning-based systems, to deal with the adversarial semantic noise. The adversarial training searches for the semantic noise, $\bm{N_{A}}$, that can fool deep models by maximizing the loss, while parameters of the system are updated to overcome the impacts of $ \bm{N_{A}} $. We denote $ \mathcal{L} (\cdot) $ as the loss function for adversarial training, which could be set as $L_{total}$ or part of $L_{total}$. An adversarial training process is typically formulated as
\begin{equation}\label{adv_opti}
 \min_{\bm{\varphi}, \bm{\eta}, \bm{\zeta}, \bm{\delta}} \mathbb{E}_{(\bm{S}, \bm{\hat{S}})\in \bm{D}}[{\max_{\bm{N_{A}} \in  \boldsymbol N}  \mathcal{L}(\bm{X_{embed}} + \bm{N_{A}}, \bm{S}, \bm{\hat{S}}; \bm{\varphi}, \bm{\eta}, \bm{\zeta}, \bm{\delta})}],
\end{equation}
where $\bm{D}$, $\boldsymbol N$ are the train set and the distribution space of $\bm{N_{A}}$. The critical problem of adversarial training is finding the most sensitive semantic noise for the system. The adversarial semantic noise, $\bm{N_{A}}$, can be explored by the FGM. The FGM generates $\bm{N_{A}}$ by
\begin{equation}\label{FGM_1}
\bm{N_{A}} = \epsilon \cdot \frac{\nabla_{\bm{X_{embed}}} \mathcal{L}(\bm{S}, \bm{\hat{S}}; \bm{\varphi}, \bm{\eta}, \bm{\zeta}, \bm{\delta})}{\left\| \nabla_{\bm{X_{embed}}} \mathcal{L}(\bm{S}, \bm{\hat{S}}; \bm{\varphi}, \bm{\eta}, \bm{\zeta}, \bm{\delta}) \right\|_2},
\end{equation}
where $\epsilon$ is the normalization factor. The adversarial semantic noise can be obtained using back propagation. 

The adversarial semantic noise added to the embedding layer could affect the semantic fidelity of semantic communication systems, the adversarial training is able to enhance the robustness of the semantic communication system. In this paper, the FGM method is adopted to improve the robustness of the semantic communication system. The training set is augmented with adversarial examples $\bm{X_{embed} + N_A}$ that are crafted by FGM and the system model is trained to against the adversarial noise.

\subsection{Performance Metrics}

Compared with conventional communication systems, metrics, such as bit-error rate and symbol-error rate, are unable to measure the performance of semantic communication systems well. For semantic communications, it is necessary to consider whether there is a semantic gap between the shared text and the received text. Hence, we use the BLEU score~\cite{bleu} and the BERT SCORE~\cite{bertscore} to describe the performance of the system comprehensively, which are detailed in the following.

\subsubsection{BLEU Score}

The BLEU utilizes the n-gram matching criterion to evaluate the quality of a received text. We denote $C_k$ as the number of the k-th word for the n-gram text, $W_n$ as the weight of the n-gram precision, and $BP$ as the penalty index. The BLEU score is obtained as follows.

\begin{small}
    \begin{equation}\label{bleu}
        % BLEU = BP \times \exp{\sum_{n=1}^{N}W_n P_n}
        BLEU = BP \times \exp({\sum_{n=1}^{N}W_n \frac{\sum_{i}\sum_{k}\min{(C_k(R_i)), C_k(T_i))}}{\sum_{i}\sum_{k}C_k(R_i)}}.
    %  \nonumber
    \end{equation}
\end{small}

Particularly, $BP$ is defined as
%where $P_n$ is the modified n-gram precision, $BP$ is the penalty index, and $W_n$ is the weight of the n-gram precision. The following can obtain $P_n$ and $BP$.
%\begin{equation}
%P_n = \frac{\sum_{i}\sum_{k}\min{(Count_k(R_i), Count_k(T_i))}}{\sum_{i}\sum_{k}Count_k(R_i)}
%\end{equation}
% where $Count_k$ is the number of the k-th word for the N-gram text.
\begin{equation}
BP=\left\{
\begin{array}{rcl}
1, & & l_R > l_T,\\
e^{1-\frac{l_R}{l_T}}, & & l_R < l_T,
\end{array} \right.
\end{equation}
where $l_R$ is the length of the received text, and $l_T$ is the length of the transmitted text. The value of the BLEU score is between 0 and 1, and the higher score implies greater sentence similarity. The BLEU score is effective but it only evaluates the similarity in the literal variation, rather than the semantic difference. Therefore, we also use BERT SCORE as the metric to depict the semantic similarity between two sentences.

\subsubsection{BERT SCORE}

The BERT SCORE obtains the semantic similarity from a similarity matrix and applies different weights to words according to their corresponding semantic importance. Thus, the semantic similarity evaluated by BERT SCORE correlates well with human judgments.

We assume that the corresponding BERT representation vector of transmitted text $\bm{S}$ is $\left< \bm{T_1}, \bm{T_2}, \dots, \bm{T_n} \right>$, and representation vector of the received text $\bm{\hat{S}}$ is $ \left< \bm{R_1}, \bm{R_2}, \dots, \bm{R_m} \right> $. The importance weight function $idf (\cdot) $ can be obtained by
\begin{equation}
idf(x) = -\log \frac{1}{M} \sum_{1}^{M} \mathbb I (x \in \bm{R^{(i)}}),
\end{equation}
where $\{ \bm{R^{ (0) }}, \bm{R^{ (1) }}, \dots, \bm{R^{ (M) }} \}$ is the test corpus. 

The precision of the BERT SCORE between the transmitted text and the received text can be obtained as
\begin{equation}
    P_{BERT} = \frac{\sum_{r_i \in \bm{\hat{S}}} idf(r_i) \max_{t_i \in \bm{S}} \bm{T_i ^\mathrm{T}} \bm{R_i}}{\sum_{r_i \in \bm{\hat{S}}} idf(r_i)}. 
\end{equation}

Then, the BERT SCORE is scaled to a larger interval using the following transformation to make it more readable by
\begin{equation}
    \hat{P}_{BERT} = \frac{P_{BERT}-b}{1-b},
\end{equation}
where $b$ is a scale factor. The rescaled BERT SCORE is between -1 and 1, and a higher score implies greater similarity between the compared sentence pair.

\section{Numerical Results}

In this section, we conduct experiments to evaluate our developed R-DeepSC under various forms of semantic noise. 

\subsection{Corpus and Baseline Models}

Europarl~\cite{Europarl} has been adopted as our data set, which is based on proceedings of the European Parliament in 11 languages. We have selected Europarl in English, which contains 98, 751 sentences, as the transmitted corpus. 4 kinds of errors have been added to each sentence in this corpus randomly, including replacement, random mask, insertion, and verb errors. 

This paper chooses two systems as comparisons. One is the DeepSC based on deep learning, and another one is a conventional communication system that uses Huffman codes for source coding, the Reed-Solomon (RS) codes for channel coding, and 64-QAM for modulation.

We evaluate system performance under different channel environments, including additive white Gaussian noise (AWGN) channels, and Rayleigh fading channels. R-DeepSC is robust to semantic noise by conducting adversarial training with the FGM and utilizing a calibrated self-attention mechanism.

\subsection{Experimental Results}

\begin{figure}[htp]
	\centering
	\includegraphics[scale=0.25]{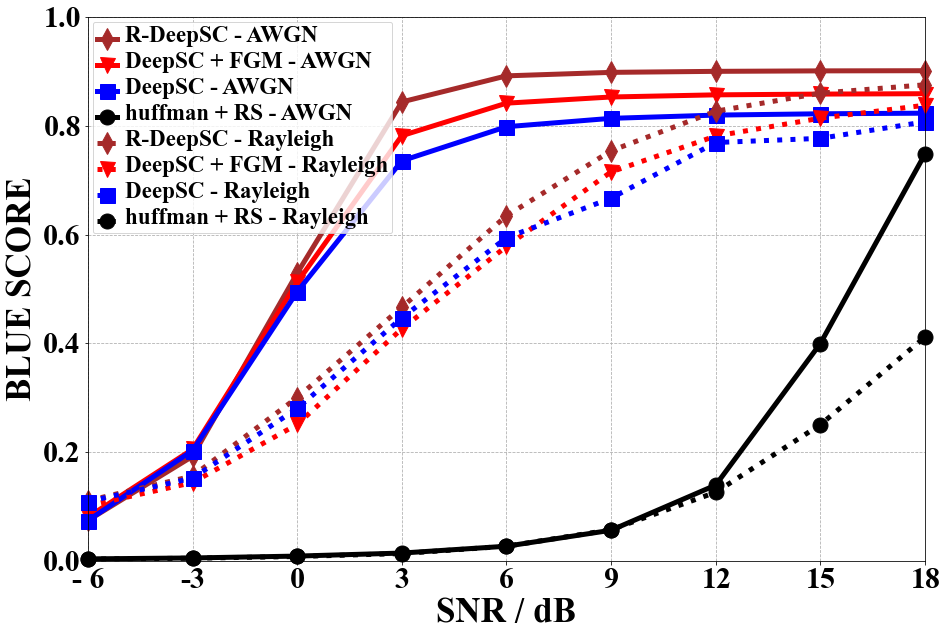}
	\caption{BLEU score versus SNR.}
	\label{fig:bleu_0.2}
\end{figure}

Fig.~\ref{fig:bleu_0.2} shows BLEU scores of systems when the corpus contains 20\% literal errors for each sentence. It can be seen that when SNR is below 12 dB, the conventional communication system using Huffman coding and RS coding has a great performance decline in terms of BLEU and BERT SCORE. When SNR increases to 18 dB, although the BLEU score of the conventional system gradually increases to nearly 80\%, there is still a non-negligible performance gap between the conventional approach and deep learning-based methods, such as the R-DeepSC and DeepSC.

The conventional system is unable to correct semantic errors due to the lack of semantic perception, so the BLEU score can hardly exceed 80\%. While semantic communication systems extract semantic information, they can correct erroneous text to some extent. Among these semantic communication systems, our proposed R-DeepSC achieves superior performance under different SNRs. These results demonstrate that the semantic communication system can mitigate semantic distortion during transmission, while the R-DeepSC outperforms other methods.

\begin{figure}[htp]
	\centering
	\includegraphics[scale=0.25]{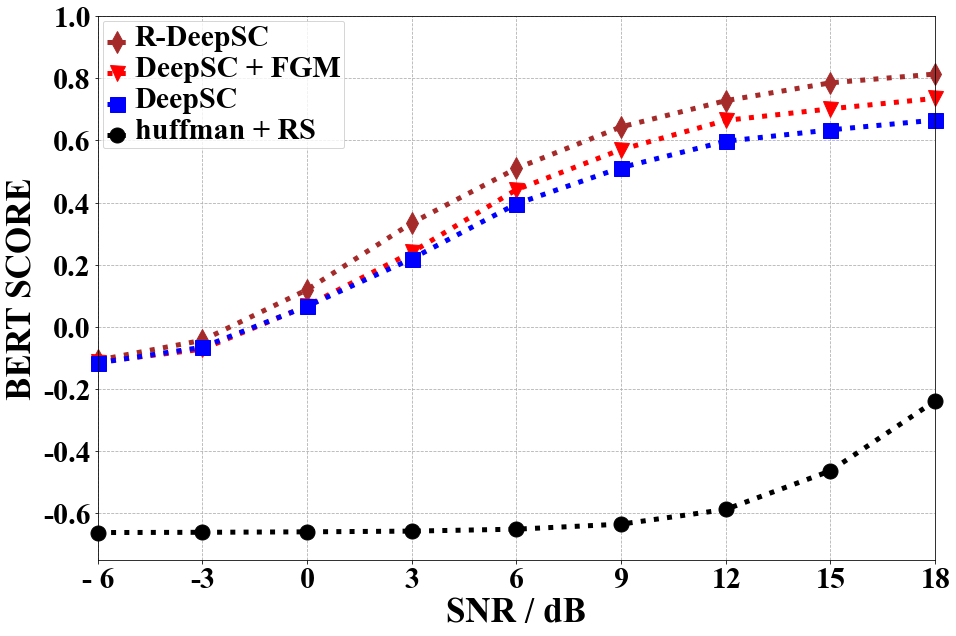}
	\caption{BLEU score versus SNR under Rayleigh fading.}
	\label{fig:bert_0.2}
\end{figure}

In addition, the effectiveness of the FGM is validated. For BLEU score, DeepSC trained with the FGM (labelled as DeepSC+FGM) performs better. Meanwhile, as shown in Fig.~\ref{fig:bert_0.2}, if we measure the system performance with the BERT SCORE, which calculates the semantic similarity, the DeepSC trained with the FGM shows the same tendency in Rayleigh fading channels. When SNR is lower than 0 dB, the FGM can hardly promote the system's performance because the distortion is too severe. As SNR increases, the FGM can improve the semantic fidelity of decoded texts effectively.

\begin{figure*}[htbp]
    \centering
    \subfigure[BLEU score versus semantic noise ratio]{
        \includegraphics[width=8cm]{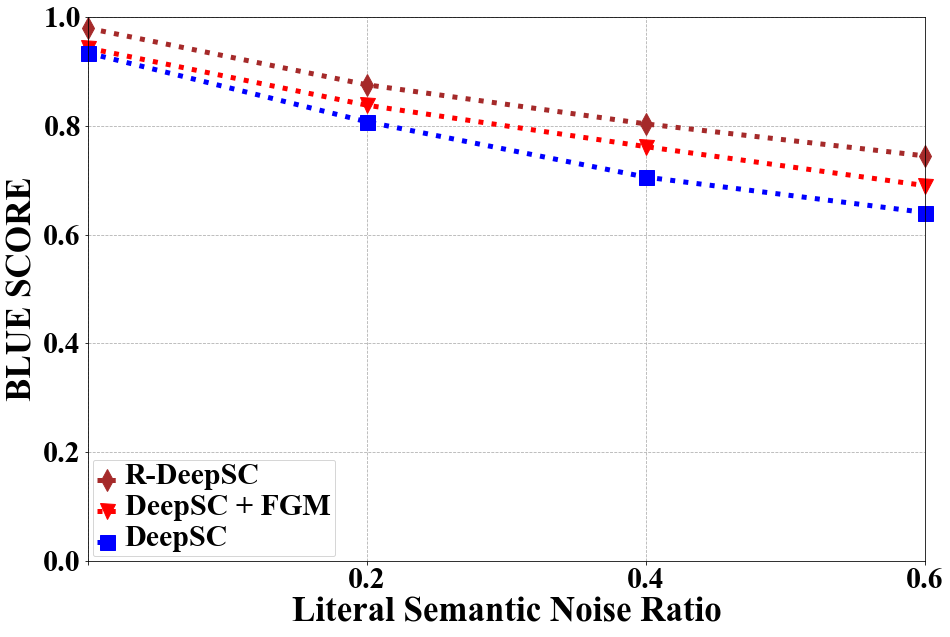}
        \label{fig:bleu}
        }
    \quad
    \subfigure[BERT SCORE versus semantic noise ratio]{
        \includegraphics[width=8cm]{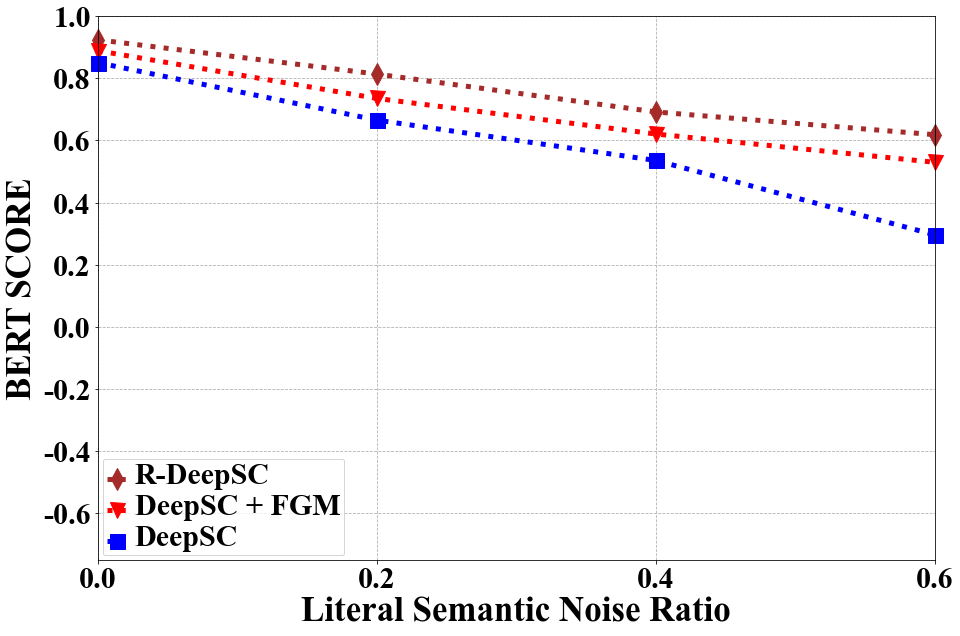}
        \label{fig:bert}
        }
    \quad
    \caption{Performance of systems trained with the corpus that contain different levels of literal errors under Rayleigh fading.}
    \label{fig:performace_2}
\end{figure*}

Moreover, we conducted experiments in scenarios with different levels of literal semantic noise. Fig.~\ref{fig:performace_2} shows the results trained under different literal semantic noise ratios. Fig.~\ref{fig:bleu} presents that although the semantic fidelity obtained by the semantic communication system decreases when the literal semantic noise ratio increases, our proposed R-DeepSC yields remarkable performance under Rayleigh fading channels. At the same time, Fig.~\ref{fig:bert} shows that the semantic fidelity of R-DeepSC decays more slowly as the proportion of the literal semantic noise in corpus increases to 60\%, which indicates that our method is indeed semantic noise-robust.

% \begin{figure*}[htbp]
%     \centering
%     \subfigure[BLEU score versus SNR]{
%         \includegraphics[width=8cm]{figure/0.2_BLEU.png}
%         \label{fig:bleu_0.2}
%         }
%     \quad
%     \subfigure[BERT SCOR versus SNR under Rayleigh fading]{
%         \includegraphics[width=8cm]{figure/0.2_Bert.png}
%         \label{fig:bert_0.2}
%         }
%     \quad
%     \caption{Performance under different SNRs on a corpus containing 20\% errors.} %Solid lines and dashed lines represent performance under AWGN and Rayleigh fading, respectively.}
%     \label{fig:performance_1}
% \end{figure*}

An example of the decoded text is shown in Table~\ref{table_example}. About 20\% words of the sentence are modified by literal errors that incur semantic distortion. We can see that most errors in texts can be corrected after being transmitted by R-DeepSC and the original semantics of the text are restored. The literal semantic noise, such as verb errors, and insertions, can be eliminated effectively, while some trivial information is filtered. For example, the name "Emma Bonino" is interpreted as "Bonino", but this modification can hardly affect its underlying meaning. 

In summary, the proposed R-DeepSC, which yields remarkable performance compared with other systems, can effectively correct semantic distortions caused by modifications in texts and adversarial noise. This performance improvement not only comes from the developed architecture and calibrated self-attention mechanism of R-DeepSC, but also from taking advantage of the adversarial training.

\begin{table*}[!h]
    \renewcommand{\arraystretch}{1.0}
    \caption{Decoding Results for Samples Containing 20\% Literal Semantic Noise}
    \label{table_example}
    \centering
    \begin{tabular}{|c|c|}
    
        % \hline
        % Text with \textcolor{red}{Errors}
        % & The president cut \textcolor{red}{later} the speaker off. \\ 
        % \hline
        % Correct Text 
        % & The president cut the speaker off.\\
        % \hline
        % Transmitted by R-DeepSC
        % & The president cut the speaker off.\\
        % \hline
        % Transmitted by  DeepSC
        % & The president cut off the speaker off.\\

        \hline
        Text with \textcolor{red}{Errors}
        & \begin{tabular}[c]{@{}l@{}} 
        It \textcolor{red}{give} \textcolor{red}{*} great pleasure to welcome \textcolor{red}{could} Emma Bonino the italian \textcolor{red}{ministered} \\ for european policies and international trade to the house today.\end{tabular} \\ 
        \hline
        Correct Text 
        & \begin{tabular}[c]{@{}l@{}} It gives me great pleasure to welcome Emma Bonino the italian minister \\ for european policies and international trade to the house today.\end{tabular}\\
        \hline
        Transmitted by R-DeepSC
        & \begin{tabular}[c]{@{}l@{}}
        It gives me great pleasure to welcome to Bonino the italian minister \\for european policies and international trade to the house today.\end{tabular}\\
        \hline
        Transmitted by DeepSC
        & \begin{tabular}[c]{@{}l@{}} We have been taken to make a fewd the rapporteur for the \\ european union and the european union to the european union.\end{tabular} \\
        \hline
        
        % Text with \textcolor{red}{Errors}
        % & it \textcolor{red}{been} vital to close the legal loopholes that allow \textcolor{red}{day} these \textcolor{red}{*} businesses to operate. \\ 
        % \hline
        % Correct Text 
        % & it is vital to close the legal loopholes that allow these fraudulent businesses to operate.\\
        % \hline
        % Decoded Text by R-DeepSC
        % & it is vital to close the legal loopholes that allow all these national businesses to operate.\\
        % \hline
        % Decoded Text by DeepSC
        % & we have to the eu to the eu has not not not not to the people to be.\\
        % \hline
        
    \end{tabular}
\end{table*}

\section{Conclusion}

In this paper, we have proposed a robust semantic communication system, which combats different forms of semantic noise and improves the robustness under various wireless environments. In particular, we have elaborated on literal semantic noise and adversarial semantic noise in semantic communication systems. For the literal semantic noise, we have developed a novel semantic encoder architecture and calibrated self-attention scheme that leverages the semantic information extracted by the semantic encoder to correct literal errors. Experiments show the effectiveness of our proposed R-DeepSC when the corpus is erroneous. For the adversarial semantic noise, we have adopted the adversarial training method to find perturbations that disturb the semantic communication system mostly and train our system to resist these perturbations. The experimental results demonstrate that eliminating adversarial semantic noise can improve the performance of semantic communication systems under different SNRs.

\section*{Acknowledgment}

This work was supported by the National Key R\&D Program of China (2018YFB1800804), the National Natural Science Foundation of China (NSFC 61925105, 61801260, 62101307), and the fellowship of China National Postdoctoral Program for Innovative Talents (BX20200194). This work was also supported by Tsinghua University-China Mobile Communications Group Co., Ltd. Joint Institute and Shanghai Municipal Science and Technology Major Project (Grant No.2018SHZDZX04).

\bibliographystyle{IEEEtran}
\bibliography{reference.bib}

%\end{IEEEbiography}

\end{document}